\documentclass[iop]{emulateapj}
\usepackage[]{graphicx}

\def\pasa{Publ. Astron. Soc. Australia}
\def\mearth{M$_{\earth}$}

\shortauthors{KERR ET AL.}
\shorttitle{Planet Formation Around Young Pulsars}

\begin{document}

\title{Limits on Planet Formation Around Young Pulsars and Implications for Supernova Fallback Disks}

\author{
M.~Kerr\altaffilmark{1,2},
S.~Johnston\altaffilmark{1},
G.~Hobbs\altaffilmark{1},
and R.~M.~Shannon\altaffilmark{1}
}
\altaffiltext{1}{CSIRO Astronomy and Space Science, Australia Telescope National Facility, PO~Box~76, Epping NSW~1710, Australia}
\altaffiltext{2}{email: matthew.kerr@gmail.com}

\begin{abstract}
We have searched a sample of 151 young, energetic pulsars for
periodic variation in pulse time-of-arrival arising from the influence
of planetary companions.  We are sensitive to objects with masses two
orders of magnitude lower than those detectable with optical transit
timing, but we find no compelling evidence for pulsar planets.  For
the older pulsars most likely to host planets, we can rule out Mercury
analogues in one third of our sample and planets with masses
$>$0.4\,\mearth{} and periods $P_b<$1\,yr in all but 5\% of such
systems.  If pulsar planets form primarily from supernova fallback
disks, these limits imply that such disks do not form, are confined to
$<$0.1\,AU radii, are disrupted, or form planets more slowly
($>$2\,Myr) than their protoplanetary counterparts.
\end{abstract}

\keywords{pulsars:general,planets and satellites:formation}

\section{Introduction}

Somewhat surprisingly, the first extra-solar planets were detected in
orbit around the millisecond pulsar PSR~B1257$+12$ \citep{Wolszczan90}.
Since then, optical observers have honed their techniques
to the point where thousands of planets have now been detected around
virtually all stellar classes \citep[e.g.][]{Wright11,Rowe14}.
Meanwhile, in spite of potentially high sensitivity to low-mass
planets \citep{Thorsett92,Cordes93}, the radio pulsar community has
turned up only one further case, a super-Jupiter in a triple system
within a globular cluster \citep[PSR~B1620$-$26,][]{Sigurdsson03}.

Why is this so? First, a Galactic field pulsar planet must either have
survived a supernova or have formed from the debris of the explosion.
\citet{Phinney93} give a delightful review of these ``Salamander'' and
``Memnonides'' scenarios, of which we consider planet formation in a
supernova fallback disk \citep{Lin91} the most relevant mechanism.
Here, reverse shock waves generated at density transitions in the
stellar core and envelope decelerate material that falls back
toward the compact remnant \citep{Chevalier89}.  The total mass,
accretion rate, and intrinsic angular momentum depend sensitively on
both the initial stellar conditions and mixing during the explosion,
but ultimately $\sim$\,0.1\,M$_{\sun}$ of material may circularize
into a relatively long-lived ``debris'' disk \citep{Menou01}.
\citet{Wang06} observed a distinct infrared component from the
anomalous X-ray pulsar 4U\,0141$+$61, consistent with the reprocessing
of X-ray emission by a warm disk, but other direct disk searches
have failed to locate further examples \citep[e.g.][]{Wang14}.

Secondly, pulsars are far from perfect clocks.  A companion induces a
neutron star reflex motion detectable as a periodic delay in the pulse
time-of-arrival (TOA).  But pulsars suffer large amplitude, red timing
noise \citep{Shannon10} and glitches which can mask such periodic
signals.  In addition, correlations between pulse shape and timing
residuals indicate quasi-periodic state switching
\citep{Lyne10,Hobbs10b} with time-scales of months to years, mimicking
the signature of planets.  Finally, planets with very short or very
long periods can evade detection due to finite observing cadences and
data spans, respectively.

\begin{figure}
\centering
\includegraphics[angle=0,width=0.45\textwidth]{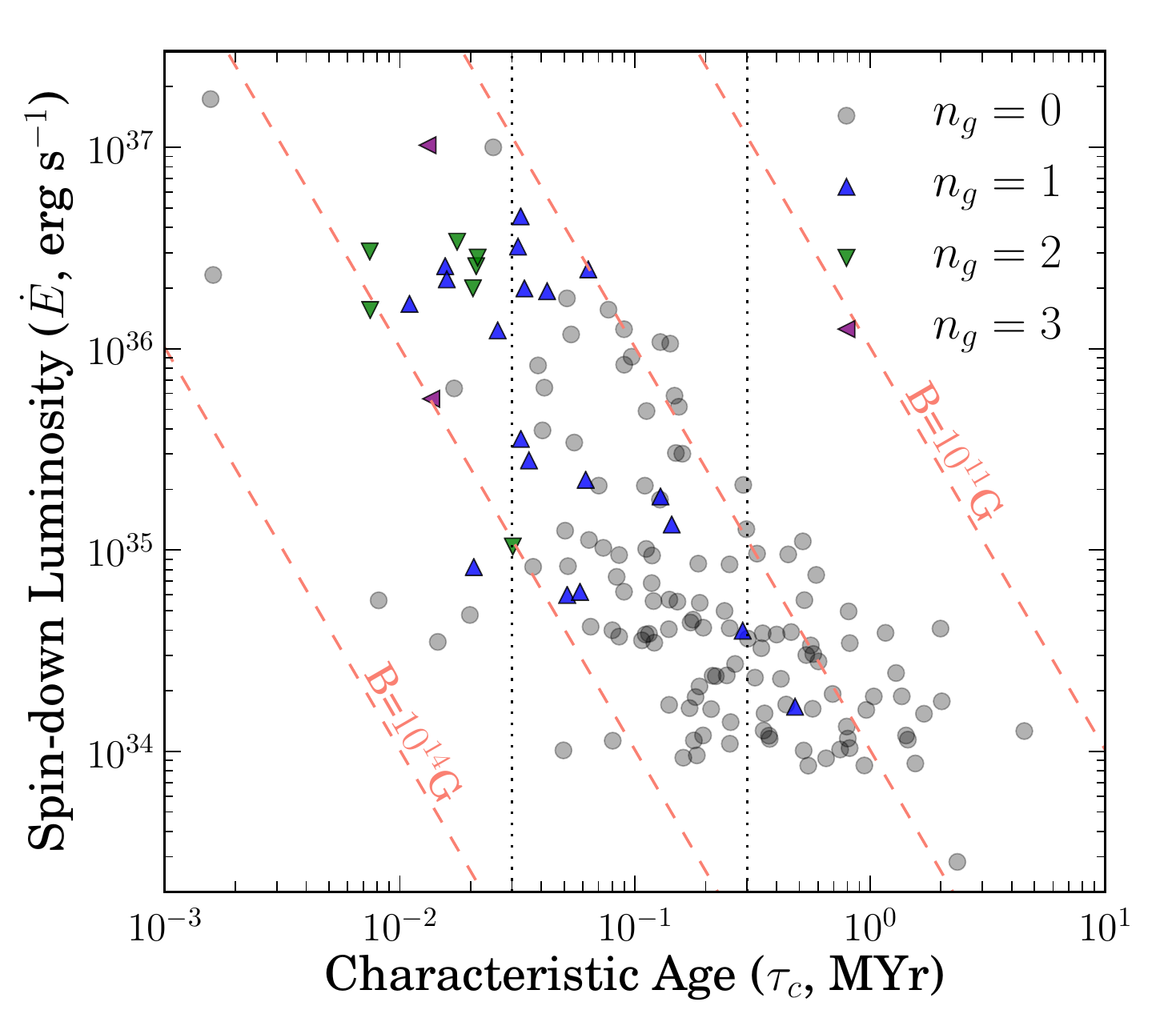}
\caption{\label{fig:edots}The spin-down power ($\dot{E}\equiv
10^{45}\dot{P}/P^3$\,erg\,s$^{-1}$) and characteristic ages
($\tau_c\equiv P/2\dot{P}$\,s) of the 151 pulsars in the sample.  The
sample divisions (see main text) in $\tau_c$ are indicated by the
vertical lines.  Pulsars
suffering large glitches ($\delta\nu/\nu>10^{-7}$) during our
observation span are indicated by coloured, distinct symbols (legend).
Lines of constant magnetic field
($B=3.2\times10^{19}(P\dot{P})^{1/2}$\,G) are shown as dashed salmon
lines.}
\end{figure}

To further investigate the paucity of pulsar planets, we have searched
for periodic signals in a sample of 151 energetic (spin-down
luminosity $\dot{E}>10^{34}$\,erg\,s$^{-1}$) pulsars timed with the
Parkes telescope in support of the \emph{Fermi} mission.  The
$\dot{E}$ and characteristic age $\tau_c$ of the sample are shown in
Figure \ref{fig:edots}.  These young pulsars tend to glitch regularly
and are also subject to strong timing noise, complicating the search
for planets with periods much longer than a year.  However, given the
angular momentum budget ($\sim$$10^{49}$\,erg\,s) of a fallback disk,
any Earth-mass planets should reside within $\sim$2\,AU of the neutron
star, making our insensitivity to long periods unimportant for
searching for planets formed in such disks.  Moreover, the age range
of our sample precisely spans the lifespan of an active disk and
subsequent planet formation.

Below, we present a new technique (\S\ref{sec:data}) for jointly
modelling stochastic timing noise with other timing parameters which
improves the robustness of detections of and upper limits on periodic
signals.  We derive upper limits (\S\ref{sec:ul}) for a range of
periods for each member of our sample and present the combined
constraints on the young pulsar population.  We interpret the limits
and their implications for supernova fallback debris disks in
\S\ref{sec:discussion_disks} and we summarize our results in
\S\ref{sec:summary}.

\section{Data Analysis}
\label{sec:data}

Our data comprise monthly timing observations with the Parkes
telescope of 151 pulsars, largely identical to those given by
\citet{Weltevrede10}.  The bulk of these observations are carried out
with the 20\,cm multi-beam receiver, with 256\,MHz of bandwidth
centred at 1369\,MHz digitally polyphase filterbanked into 1024
channels and folded in real-time into 1024 phase bins.  Observation
lengths $t_{\mathrm{obs}}$ range from 2--20 minutes depending on the
flux density and sharpness of the pulse profile.  Data typically span
the seven years from 2007 June to 2014 Oct, and within this interval
we supplement our observations with archival data where available.

\begin{figure}
\includegraphics[angle=0,width=0.45\textwidth]{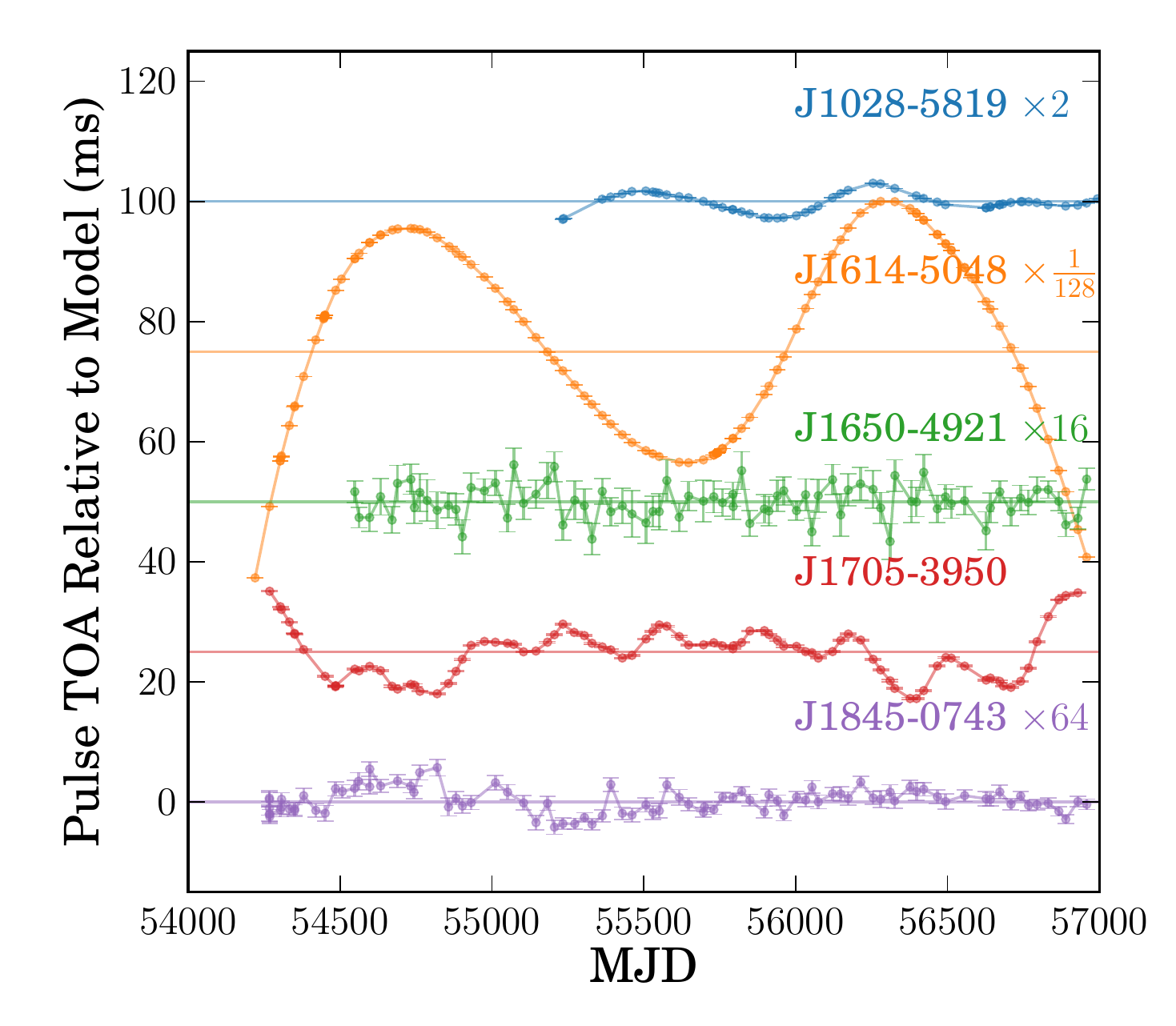}
\caption{\label{fig:residuals}The deviation of TOAs of five pulsars to
a simple spin-down model including $\ddot{\nu}$ to reduce
low-frequency timing noise.  The pulsars are chosen to illustrate the
range of red and white noise in the sample, and residuals have been
scaled by the indicated amount to facilitate comparison.}
\end{figure}

For each pulsar, we have determined the best timing model, typically
comprising spin-down terms $\nu$ and $\dot{\nu}$, the pulsar position,
and, where required, proper motion and glitch parameters.  A
representative sample of the timing residuals to such a model
(including a $\ddot{\nu}$ term) appears in Figure
\ref{fig:residuals}.
The pulsars are strongly affected by red ``spin'' noise, and to
estimate these parameters robustly requires modelling the resulting
covariance between TOA measurements.  We assume the red noise process
is widesense stationary and model its power spectral density as
\begin{equation}
P(f) = A_0 \left[1 + (f/f_c)^2\right]^{-\alpha/2}.
\end{equation}
Here, $f_c$ allows a possible low-frequency cutoff to the timing
noise, while $\alpha$ describes its shape. By the Wiener-Khinchin
theorem, $P(f)$ is the Fourier transform of $C(\tau)$, the covariance
between TOAs separated in time by an interval $\tau$.  To this
$C(\tau)$ we add the white noise model, a diagonal covariance matrix
comprising measurement uncertainty and ``jitter'' noise
\citep{Shannon14}, which we model as
$\sigma_{j}^2=\beta/t_{\mathrm{obs,j}}$.  The complete covariance
matrix $C$ can be written as a Cholesky decomposition, $C = LL^*$,
with $L$ a lower triangular matrix.  If the noise model is accurate,
$L$ transforms the TOAs into unit variance normal random variables
\citep{Coles11}.  The timing noise and jitter parameters, together
with the pulsar timing parameters, form a complete model, and with the
Cholesky decomposition we can efficiently evaluate its log likelihood.
We apply Monte Carlo Markov Chain methods
\citep[MCMC,][]{Foreman-Mackey13} to determine its maximum and shape,
yielding estimators for the parameters and their statistical
uncertainty.  We generally find this approach results in an acceptable
fit, as evidenced by the successful whitening of the residuals,
justifying our adoption of a power-law red noise model.


To characterize the reflex motion due to a single planet, we add to
the above model the five Keplerian parameters (orbital period $P_b$,
projected semi-major axis $a$, epoch of periastron $T_0$, eccentricity
$e$, and longitude of ascending node $\Omega$) that fully define the
orbit.  We restrict the parameter space to $0<e<0.3$ (with uniform
prior) and $56<P_b\,(\mathrm{d})<5000$.   This is roughly the range of
eccentricity expected from oligarchical planet formation scattering
(see \S\ref{sec:discussion_disks}).  The lower $P_b$ bound is
critically sampled with our monthly observing cadence, while the upper
bound is roughly twice the data span.  This is a substantial parameter
space, and to ensure we explore it fully, we divide it into three
subspaces: $56<P_b\,(\mathrm{d})<341$, $341<P_b\,(\mathrm{d})<393$,
and $393<P_b\,(\mathrm{d})<5000$, a division motivated below.
Following a lengthy burn-in period for the Markov chains to ``forget''
their initial conditions, we draw $1.2\times10^6$ samples from the
first subspace and $6\times10^5$ samples from the latter two to form
our final MCMC sample.

To search for planets, we computed the change in the log likelihood
$\delta\log\mathcal{L}$ between the null (no planet) model and the
best-fitting planet model over $P_b$ bins.  The value
$\delta\log\mathcal{L}$ for a given planet mass depends strongly on
the level of white and red noise, but with simulations we established
$\delta\log\mathcal{L}\approx12$ as a reasonable universal threshold
for a significant detection.

Restricting attention to candidates with $P_b<2$\,yr, we found
significant modulation in five pulsars,with the timing residuals of
one, PSR~J1705$-$3950, appearing in Figure \ref{fig:residuals}.  There
are two strong arguments suggesting these are not pulsar planets.
First, the implied masses are as great as 6\,\mearth{}, and
such objects would be easily detected, but have not been, in samples
including older pulsars with little timing noise
\citep[e.g.][]{Hobbs10b}.  Second, three show evidence for a
significant second harmonic, similar to the well-known state switching
PSR~B1828$-$11 \citep{Stairs00,Lyne10}, inviting a magnetospheric
interpretation.  If these harmonics are instead interpreted as
two-planet systems, the resulting 2:1 mean motion resonance is
dynamically unstable for the mass ratios \citep{Beauge03}.  We
will present a detailed analysis and interpretation of these
modulations in a future work, but emphasize here the key conclusion,
that these modulations are unlikely to represent \textit{bona fide}
planets, and that the small number of candidates ensures that the
upper limits we compute below remain valid.

\section{Upper Limits on Planet Masses}
\label{sec:ul}

The Keplerian parameter $a$ is related to the companion mass by the
mass function:
\begin{equation}
\label{eq:kepler}
\frac{m_c\,\sin i}{M_{\earth}}=\frac{a}{1.50\times10^{-3}\,\mathrm{lt\,s}}\left(\frac{M_*}{M_{\sun}}\,\frac{1\,\mathrm{yr}}{P_b}\right)^{2/3},
\end{equation}
where we have assumed $m_c\ll M_*$, with $M_*$ the neutron star mass.
While masses of isolated neutron stars are generally unknown,
measurements from unrecycled binary systems suggest the Chandrasekhar
mass $M_*=1.4\,M_{\sun}$ is representative \citep[e.g.][]{Lattimer12}.
If the plane of the putative binary is randomly orientated, i.e.
$p(i)=\sin i$, then the mean and median values of $\sin i$ are
$\pi/4=0.785$ and 0.866 respectively.  We report projected mass values
$m_c\,\sin i$, so physical masses corresponding to our upper limits
will typically be 15\% greater.

\begin{figure}
\includegraphics[angle=0,width=0.45\textwidth]{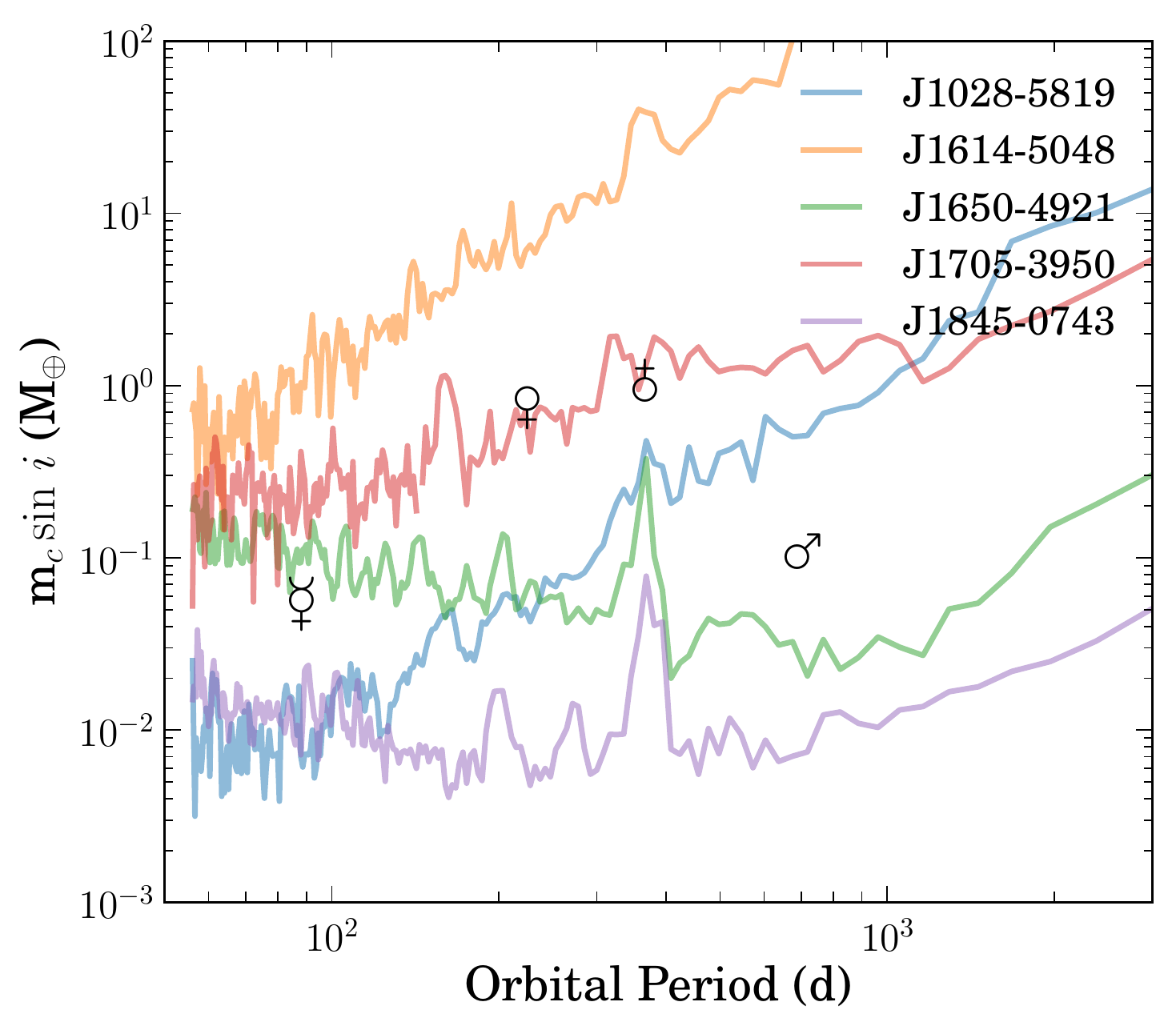}
\caption{\label{fig:ul_examples}The 90\% containment interval of the
posterior distribution for companion mass, for five representative
pulsars, as a function of companion period.  Symbols indicate the
masses and orbital period for inner solar system planets Mercury,
Venus, Earth, and Mars.}
\end{figure}

Several examples of the resulting planet limits, as a function of
$P_b$, are shown in Figure \ref{fig:ul_examples}.  As noted by
\citet{Thorsett92}, the general features of these spectra reflect
simple physics and the relative strength of the white and red noise.
The high-frequency limit is set by white noise in the data, in turn
set by the observing cadence and the TOA precision.  In the absence of
red noise, this limit scales as $P_b^{-2/3}$ following Equation
\ref{eq:kepler}.  The sinusoid induced by planets with
$P_b\approx1$\,yr can be absorbed by shifting the apparent pulsar
position, resulting in a dramatically reduced sensitivity.  However,
because we generalize our model to include $0<e<0.3$, sensitivity at
these time-scales is actually \emph{improved} relative to the circular
orbit case.  Finally, the partial sinusoid of a long-period planet is
similar to a parabola and is degenerate with $\nu$, $\dot{\nu}$, and
the realization of timing noise, admitting large values of $a$.  Our
division into three subspaces reflects this behaviour.

Figure \ref{fig:ul_examples} shows the range of typical cases.
PSRs~J1028$-$5819 and J1614$-$5048 are both affected by red noise and
their sensitivity decreases monotonically with increasing $P_b$,
but the low white noise level for J1028$-$5819 allows good sensitivity
to short-period objects like Mercury.  PSR~J1016$-$5819 has a
relatively high white noise level but very little red noise, and the
favorable scaling of sensitivity with $P_b$ yields good sensitivity
to objects with $P_b>1$\,yr, like Mars.  Finally, J1845$-$0743 offers
low noise at all frequencies and provides good sensitivity to every
inner solar system analogue.

\begin{figure}
\includegraphics[angle=0,width=0.45\textwidth]{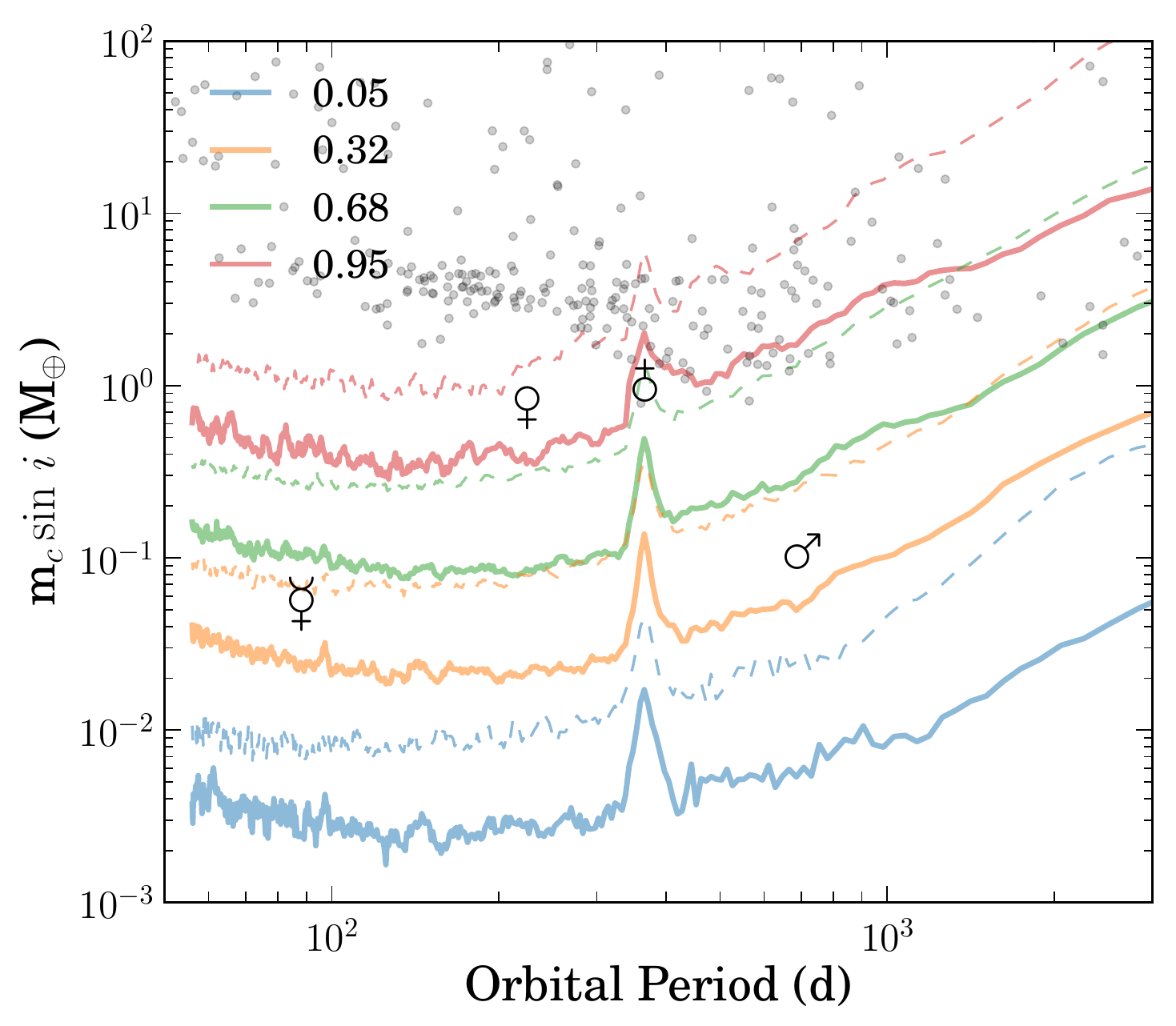}
\caption{\label{fig:ul_population}Limits on planet masses
for pulsars with $\tau_c>0.3$\,Myr (solid lines) and
$0.03<\tau_c\,/\,\mathrm{Myr}<0.3$
(dashed, faint lines).  The lines indicate the fraction of the
posterior distribution of companion mass, for each period bin,
contained below.  The grey points correspond to confirmed exoplanets,
primarily from Kepler, and are taken from the Exoplanet Database
\citep{Wright11}.}
\end{figure}

We can extract a posterior probability distribution for the entire
population by assuming each pulsar is equally likely to host a
planet.  Then the posterior distribution of companion mass in a period
bin is simply the sum of the individual distributions.  Because we do
in fact expect different pulsar ages to correspond to different stages
of disk activity and planet formation, we have divided our sample into
three bins of characteristic age ($\tau_c$ in Myr): $\tau_c>0.3$ (47
pulsars); $0.03<\tau_c < 0.3$ (84); and $\tau_c<0.03$ (20).  Our
primary result appears in Figure \ref{fig:ul_population}, where we
have constructed posterior distributions for the first two $\tau_c$
bins.  The curves show the companion mass below which the indicated
fraction of the posterior distribution is contained.  They can be
interpreted as limits on the occurrence of planets of a given mass:
there is only a 5\% chance a planet as massive as those indicated by
the uppermost curve could reside in our sample, etc.  For
completeness, we performed a similar analysis using a frequentist
method (profile likelihood) but assuming circular orbits, and we
obtained similar results when considering the 95\% confidence upper
limits of the population.

The oldest pulsars provide the tighest constraints, and these are the
most likely to harbor planets, having had sufficient time for disk
evolution and planet formation.  In this population, we conclude that
planets of 0.1--0.2\,\mearth{} and $P_b<$1\,yr are absent from at
least two thirds of the sample, and planets with M$>0.6$\,\mearth{}
can occur in no more than 5\% of systems.  Consequently, planet
formation within $\approx$1.4\,AU of neutron stars is at best a rare
phenomenon.

Younger systems ($0.03<\tau_c < 0.3$) provide similar but somewhat
poorer constraints: planets exceeding 0.3\,\mearth{} are absent from
at least 2/3 of our sample, and planets of mass 1--1.5\,\mearth{} must
be rare, present in no more than 5\% of cases.  The youngest systems
($\tau_c<0.03$) have low sensitivity due to their substantial timing
noise.  We expect such systems are also too young for planet formation
(see \S\ref{sec:discussion_disks}), but for completeness, we note that
our limits rule out planets of 1--2\,\mearth{} in 68\% of the very
young sample.

\section{Implications for Debris Disks}
\label{sec:discussion_disks}

Our upper limits clearly show that Earth-mass planets are rare or
absent within the first $\sim$2\,Myr of the formation of a pulsar.
What implication, if any, does this result have for the presence of
debris disks around neutron stars?  Following \citet{Phinney93} and
\citet{Miller01}, we consider it unlikely that any original planets
with periastron $<$1\,AU could have survived immersion in the stellar
envelope or remained bound after the supernova kick, and we
concentrate on \textit{in situ} planet formation.

First, we must ask: does our sample trace the young pulsar
distribution?  Our sample excludes magnetars and other obviously young
but low-$\dot{E}$ neutron stars, e.g. the weakly magnetized compact
central objects \citep{Halpern10}.  However, \citet{Watters11} and
other authors have argued that $\gamma$-ray pulsars, with properties
similar to our sample, require the bulk of Milky Way supernova
activity to account for their numbers.  We therefore believe our
selection of high-$\dot{E}$ pulsars presents a minimally biased view of
the young pulsar population, but it does not constrain disk formation
around less typical neutron stars.

Secondly: given a disk with sufficient mass and angular momentum
(radius), does planet formation occur efficiently and produce palpable
planets?  Fallback disks differ from protoplanetary disks in a key
aspect: there is no external source of material.  Thus, we appeal to a
study aimed at reproducing the properties of planets around
PSR~B1257$+$12.  \citet{Hansen09} found that disks with fallback-like
angular momenta and layered viscous accretion can produce Earth-mass
planets rapidly ($10^5$--$10^7$\,yr) at inner solar system scales and
with relatively small ($e<0.2)$ eccentricities.  Even more
encouragingly, they found that annular density profiles produced
planets of a few \mearth{} in $<$\,1\,Myr.  Such profiles might result
if dust agglomerates into grains quickly after condensation during
disk expansion, in which case the disk can deposit most of its dust at
the same radius.  In these scenarios, two or three planets with mass
1--3\,\mearth{} populate the inner AU, along with a few residual
planitesimals, and would be easily detectable in our sample.

Thus, we can say that \textit{if} such disks are common around young
neutron stars, at least some should host Earth-mass planets and we
should detect such in our sample.  Moreover, even if rapid
sedimentation does not proceed and the mass profile remains more
homogeneous, \citet{Hansen09} find the production of detectable
planets is still inevitable and merely proceeds on a 1--10\,Myr
time-scale.  The absence of Earth-mass planets in both our sample and
in larger samples including pulsars with $\tau_c>10^7$\,yr
\citep{Hobbs10b} seems to rule out such disks.

However, such disks seem to play an important r\^{o}le, at least in
the pulsar literature.  Debris disks have been advanced as triggering
magnetospheric reconfiguration \citep{Cordes08,Brook14,Kerr14},
altering the spin-down histories of pulsars \citep{Menou01b,Yan12},
contributing to pulsar timing noise \citep{Shannon13b}, and even
enabling the pulsar mechanism \citep{Michel81}.  \citet{Chatterjee00},
among others, have speculated that normal pulsars accreting from a
fallback disk may provide the X-ray emission observed in anomalous
X-ray pulsars.  They may also be responsible for some cases of
peculiar supernova light curve evolution \citep{Dexter13}.

Can we save this picture by forming disks that evade our upper limits?
After an initial period of super-Eddington accretion, a fallback disk
evolves self-similarly provided it remains viscous, transporting
angular momentum and a small fraction of the material beyond the tidal
disruption radius where planet formation can occur.  The
magneto-rotational instability \citep{Balbus91} is generally thought
to provide this viscosity so long as the disk remains ionized, but as
it cools, thermal ionization instabilities result in rapid
condensation, leaving an inactive dust disk \citep{Menou01}.  (Though
we note accretion may continue via an active central ``layer'', in
which case \citet{Currie07} find the disk can reach the right scale
for planet formation.)  The final radius of such a debris disk depends
sensitively on metallicity, and high-metallicity disks generally became
inactive at small radii ($<$10$^{10}$\,cm), within the tidal disruption
radius.  Such disks could still form planets, but would first require
scattering to move material beyond the tidal disruption radius.
Moreover, this process is impeded by ablation of small bodies by the
energetic pulsar wind \citep{Miller01} and cannot proceed until the
pulsar's spin-down power $\dot{E}$ drops low enough.

On the other hand, near impact on a companion from a supernova kick
could form a debris disk with large angular momentum
\citep{Phinney93}, giving rise to planets with periods too long to be
reliably detected in our analysis.  We expect such events to be rare,
as even in tight binaries the solid angle presented by an unevolved
companion is small, and simulations suggest most binaries will have
$P_b>10$\,d \citep{Terman98}.  Our sample contains only two binary
systems, supporting the idea that most sample members did not form in
close binaries.


Finally, it may be that long-lived debris disks simply fail to form.
Stellar material may fall back onto the compact remnant but ultimately
be ejected by a propeller phase \citep{Eksi05}.  Or, the pre-supernova
conditions may be such that the specific angular momentum of the
stellar core is too low to circularize.  In this case, the explosion
energy either unbinds the entirety of the star, leaving a neutron
star, or a huge amount of material falls back, triggering a collapse
to a black hole.  Recent one-dimensional numerical simulations by
\citet{Perna14} suggest that magnetic coupling saps the specific
angular momentum of the stellar core to such a degree that only
fine-tuned explosions can produce appreciable fallback onto a neutron
star, and the resulting disk lasts only a few hundred seconds.

\section{Summary and Conclusion}
\label{sec:summary}

We have searched a large sample of young pulsars for periodic
modulation characteristic to planetary companions.  Our work is an
improvement on previous efforts \citep{Thorsett92}, as our pulsar
sample is two orders of magnitude larger and we employ sophisticated
methods to mitigate pulsar timing noise and model realistic
(noncircular) orbits.  Despite the good sensitivity to low-mass
planets we find no compelling evidence for such systems.  We argue
that such companions could have formed in debris disks within the
2\,Myr age range spanned by our sample, and their absence implies
supernova fallback disks are either rare or confined to small radii.

Whence, then, the planets of PSR~B1257$+$12?  \citet{Miller01} have
argued PSR~B1257$+$12 was born with its present 6\,ms spin
period and an intrinsically weak magnetic field.  Such a weak field
may point to atypically weak magnetic coupling in the progenitor
stellar core, evading the constraints of \citet{Perna14}.  If this is
so, isolated millisecond pulsars with modest spin periods may make
excellent targets for direct debris disk searches.


\acknowledgements
The Parkes radio telescope is part of the Australia
Telescope, which is funded by the Commonwealth Government for
operation as a National Facility managed by CSIRO.

\bibliographystyle{mn2e}


\end{document}